\documentstyle[erice97,11pt,psfig]{article}

\begin{document}
\hfill{hep-ph/9712238}\\ 
%
\title{An Updated Analysis on Atmospheric Neutrinos
\footnote{Talk given by H. Nunokawa 
(e-mail: nunokawa@flamenco.ific.uv.es) at Erice School on Nucl. 
Physics, 19th course "Neutrinos in Astro, Particle and Nuclear Physics", 
Erice, Italy, September 1997, to appear in the proceedings. 
}}
\author{M. C. Gonzalez-Garcia$^{1.2}$, H. Nunokawa$^1$, 
O. Peres$^1$,\\T. Stanev$^3$ and J. W. F. Valle$^1$\\ 
}
\address{
\vskip .5cm
$^1$\it Instituto de F\'{\i}sica Corpuscular - C.S.I.C.\\
Departament de F\'{\i}sica Te\`orica, Universitat de Val\`encia\\
46100 Burjassot, Val\`encia, Spain
\vskip .2cm
{$^2$\it Instituto de F\'{\i}sica Te\'orica, 
             Universidade Estadual Paulista \\   
             Rua Pamplona 145,
             01405--900 S\~ao Paulo, Brazil} 
\vskip .2cm
$^3${\it Bartol Research Institute, University of Delaware \\
Newark, Delaware 197161, USA}
}

\maketitle

\abstracts{
We have reanalysed the atmospheric neutrino data 
including new results from Super-Kamiokande and 
Soudan-II experiments, under the assumption of 
two-flavor neutrino oscillation. 
We present the allowed region of oscillation parameters 
for the $\nu_\mu \rightarrow\nu_\tau$ channel.
In performing this re-analysis we also take into 
account some recent theoretical improvements in the 
flux calculations. 
}
The atmospheric neutrino anomaly \cite{review} observed in 
Kamiokande \cite{kamioka} and IMB \cite{imb}
has recently been confirmed by 
new results from Super-Kamiokande \cite{totsuka}
and Soudan-II \cite{kafka}. 
The significance of the problem is usually expressed in terms 
of the double ratio, 
$R \equiv (N_\mu/N_e)_{observed}/(N_\mu/N_e)_{predicted}$, 
where $N_\mu$ and $N_e$ stand for the number of 
$\mu$ and $e$-like events, respectively. 
All the experiments, except for Frejus \cite{frejus} and 
NUSEX \cite{nusex}, are observing the deficit in the 
double ratio, $R \simeq 0.6$. 

The most natural and simplest solution to this problem is given 
by the two-flavor neutrino oscillation, either due to 
$\nu_\mu- \nu_\tau$ or $\nu_\mu- \nu_e$ channel,  
the latter being most likely disfavoured by the recent 
result from the CHOOZ experiment \cite{chooz}. 
Under this assumption, we have re-analysed the atmospheric 
neutrino data coming from Kamiokande \cite{kamioka}, IMB \cite{imb}, 
Frejus \cite{frejus}, Nusex \cite{nusex}, as well as 
those new (preliminary) ones coming from 
Super-Kamiokande \cite{totsuka} and Soudan-II \cite{kafka}.  
In our analysis we have used one of the latest calculations 
of atmospheric neutrino flux \cite{stanev} which, in general, 
depends on energy and direction of neutrinos and solar activity,  
as well as experimental site, due to the geomagnetic effect.  
In our analysis we have also included the effect of production 
point distribution of neutrinos \cite{stanev2}, 
which should, in principle, be taken into account in 
the determination of the oscillation parameters. 

We have performed a detailed $\chi^2$ analysis, 
treating separately the $\mu$ and $e$-like data, 
but taking into account the correlation of errors. 
This is better than using the double ratio $R$, 
due to its non-Gaussian nature, 
as suggested in ref. \cite{fogli}. We present in Fig. 1 
our (preliminary) results for the $\nu_\mu- \nu_\tau$ oscillation 
channel, in (a) for each experiment and in (b) for combined 
result. We notice that the allowed region is mostly determined 
by new results from Super-Kamiokande multi-GeV binned data. 
We obtained the best fit point around $\sin^2 2\theta \sim 1$ and 
$\Delta m^2 \sim 10^{-3}$ eV$^{2}$. Let us note that the best 
fit value of $\Delta m^2$ is now lower than the one in 
the pre-Super-Kamiokande era.

More detailed description and discussion of the analysis, 
including the case for $\nu_\mu-\nu_e$ channel, will 
be given in  ref. \cite{GNPSV}. 

\psfig{file=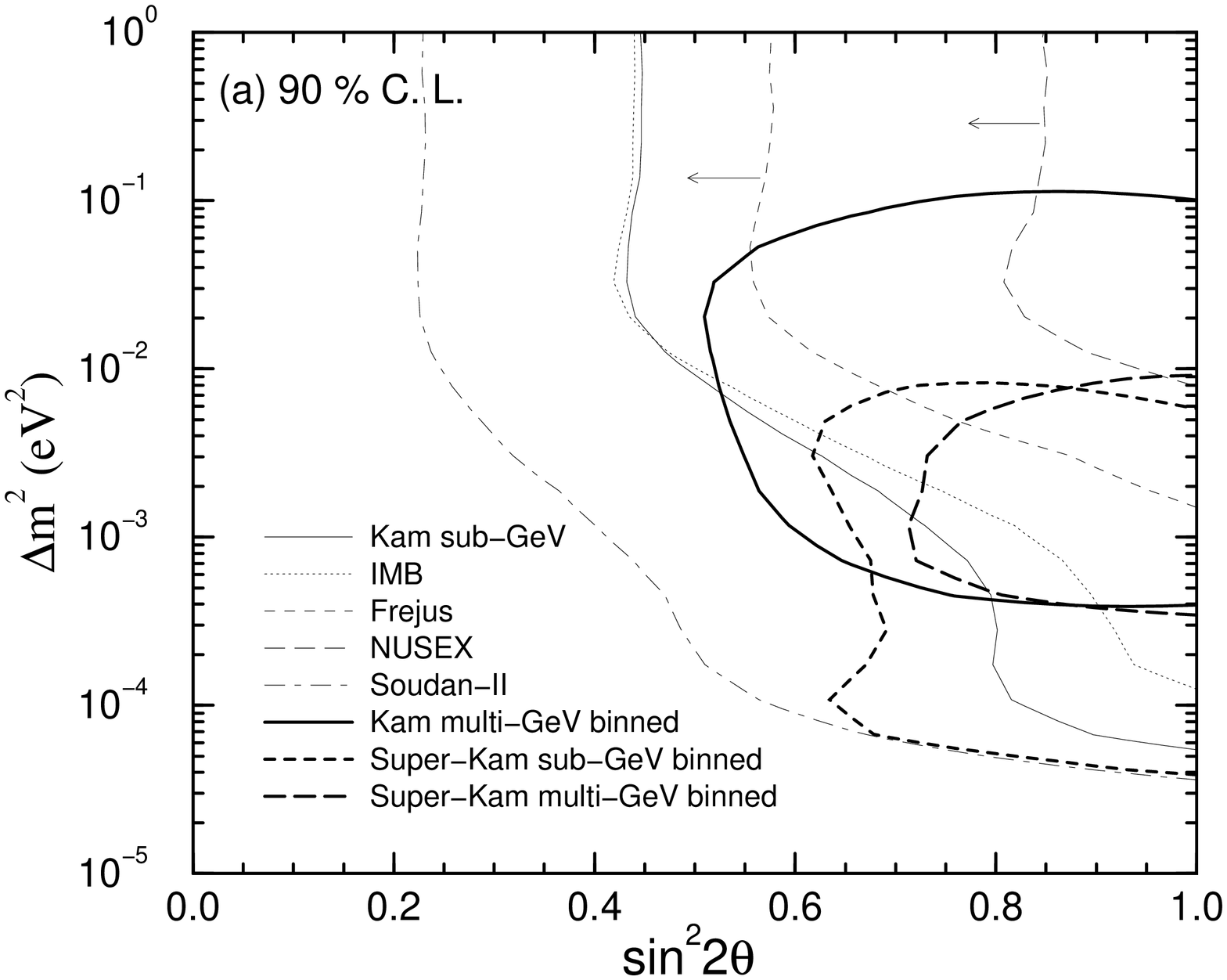,height=8.0cm,width=8.5cm,angle=-90}
\vglue -8.0cm
\hglue 7.9cm
\psfig{file=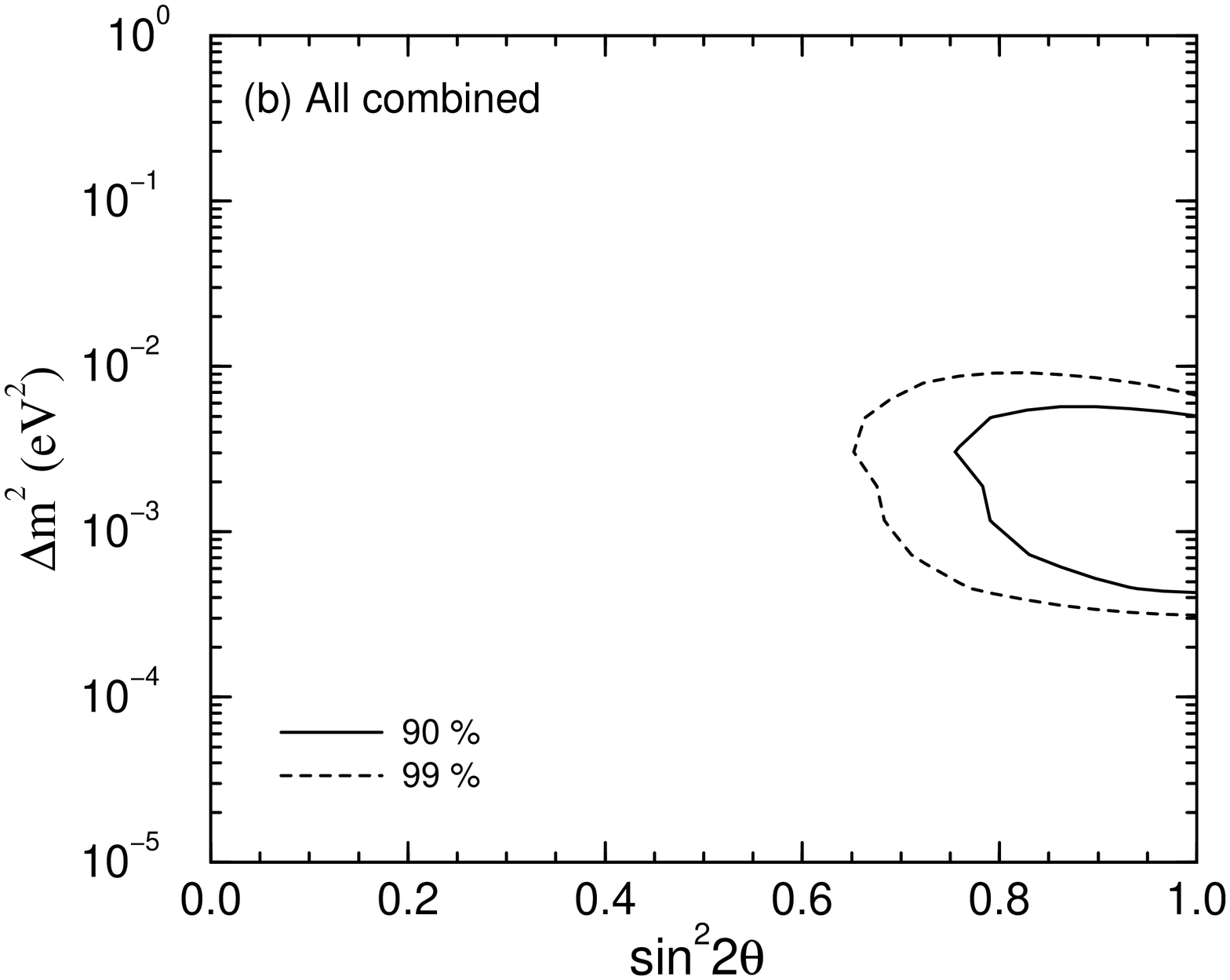,height=8.0cm,width=8.5cm,angle=-90}

\noindent
Fig. 1: In (a) we plot the individual contour for each experiment for 
90 \% C. L. and in (b) we plot the combined results for 90  and 99 
\% C. L. In (a) the regions right to the curves are allowed 
except for Frejus and NUSEX.

\end{document}